\documentclass[12pt]{iopart}

\usepackage{iopams} 
\usepackage{graphicx}
\usepackage{subfig}
\usepackage{color}
\usepackage{array}
\usepackage{multirow}
\usepackage{wrapfig}

\begin{document}

\vspace{-2cm}
\fbox{\LARGE\bf \today}

\title[]{Bidirectional Non-Markovian Exclusion Processes}

\author{Robin Jose, Chikashi Arita and Ludger Santen
 } 
\address{Department of Theoretical Physics $\&$ Center for Biophysics, Saarland University, 66123 Saarbr\"ucken, Germany}

\ead{santen@lusi.uni-sb.de}

\vspace{10pt}

\begin{abstract}\label{sect:abstract}
Bidirectional transport in (quasi) one-dimensional systems generically leads to cluster-formation and small particle currents. This kind of transport can be described by the asymmetric simple exclusion process (ASEP) with two species of particles. In this work, we consider the effect of non-Markovian site exchange times between particles. Different realizations of the exchange process can be considered: The exchange times can be assigned to the lattice bonds or each particle. In the latter case we specify additionally which of the two exchange times is executed, the earlier one (minimum rule)  or the later one (maximum rule).  In a combined numerical and analytical approach we find evidence that we recover the same asymptotic behavior as for unidirectional transport for most realizations of the exchange process. 
Differences in the asymptotic behavior of the system have been found for the minimum rule which is more efficient for fast decaying exchange time distributions.  
\end{abstract}

%
%
%
%
%

\section{Introduction}\label{sect:introduction}

One of the broadly investigated fields in non-equilibrium physics is actively driven transport.
These processes can be found in different topics such as pedestrian dynamics \cite{Chraibi2019, Gibelli2019,Helbing2004, Loevaas1994}, vehicle traffic \cite{Schadschneider2010, Chowdhury2000,Kerner2009, Helbing2004}, and intracellular transport of molecular motors along cellular filaments \cite{Burute2019,Appert-Rolland2015,Chowdhury2013, Hancock2014, Hirokawa2010,Hirokawa2005, Brown2003}.

A common tool to model active transport is a lattice gas \cite{Chou2011, Schadschneider2010}. These stochastic processes are defined in a very simple way, but lead to many interesting phenomena \cite{Mukamel2000}. Particles hop stochastically to their nearest neighbor sites on the lattice, but the hopping rates are spatially biased, and this asymmetry causes a non-vanishing flow of particles in a specific direction.
The particular case where particles are allowed to unidirectionally hop in one dimension is called the totally asymmetric exclusion process (TASEP) \cite{Derrida1998}. One of the most basic assumptions in the TASEP is that each site of the lattice is either occupied by a particle or empty. Due to this exclusion principle, hopping is prohibited if the target site is already occupied by another particle. Therefore particles behave as an obstacle for each other, 
in other words, particles themselves serve as an environment and influence motility.

In bidirectional transport, however, particles have to be transported in opposite directions.
Adaptation of the TASEP can be done by distinguishing two different species of particles with opposing directions on the same lattice. By introducing an position exchange rate, several intriguing scenarios have been reported, such as spontaneous symmetry breaking \cite{Evans1995, Arndt1998}. In case of slow position exchange interactions, the particle flux is determined by the exchange times of particles from different species similar to a defect in the unidirectional TASEP \cite{Greulich2008}.
Bidirectional TASEP models have been modified by introducing a second dimension to describe pedestrian dynamics \cite{Burstedde2001, Nowak2012} or intracellular traffic on polar filaments by adding additional lanes \cite{Klumpp2004,Ebbinghaus2009,Appert-Rolland2015}. Despite situations where symmetry is broken and the system organizes into lanes \cite{Nowak2012}, particle interactions often lead to cluster formation and is therefore a limiting factor for transport \cite{Ebbinghaus2011}.

In this paper, the focus is on active particles in confinement. Here, not only the aspect of non-equilibrium drive but also crowded and confined environments is expected to heavily influence transport processes. 
In the field of glass theory for example, a popular approach is to describe a particle trapped inside a \textit{cage}, denoting the potential created by its neighbor particles \cite{Bouchaud1992}. Such interactions can affect the waiting time distribution of particle movements as it can be seen in a trap model by Bouchaud \textit{et al.} \cite{Monthus1996}. Here, a particle falls into a trap of potential depth $E$ which is exponentially distributed and escape from it following a Poisson process with a rate depending on the energy $E$. This combination leads to algebraically (power law) distributed waiting times for particles to escape from traps. It is therefore not guaranteed that in a complex environment, properties from an exponential distribution, resulting in constant rates which are typically used in TASEP models. However, the choice of waiting time distribution can be crucial for the systems phenomena because heavy tails induce a higher statistical weight for extreme values such as for the scale free family of Pareto distributions \cite{Newman2005}.

Recent studies by Concannon \textit{et al.} \cite{Concannon2014} and Khoromskaia \textit{et al.} \cite{Khoromskaia2014} took a step forward to investigate transport behavior in the framework of unidirectional exclusion processes for non-Markovian waiting time distributions. It was found that, beside a fluid phase, the particles form condensates which are complete in space and time and hence a flux depending on the system size. This phenomenon differs from typical condensations appearing beside a stable current flowing out of clusters which is seen in models related to Markovian processes \cite{Ebbinghaus2011, Jiang2009}.


The influence of crowded environments reflected in non-Poissonian waiting times on bidirectional transport is however not fully understood. Combining the aspects of bidirectional lattice gases and non-Poissonian waiting time distributions for exchange processes, here we will investigate two-species non-Markovian TASEPs.
 
This work is organized as follows. In section \ref{sect:model}, we develop the model including three sub-versions for realizing exchange process between particles. We then show analytical estimations and simulation results for single and many particle dynamics in section \ref{sect:results} and compare them no simulation results. Finally we discuss our results in section \ref{sect:conclusion}.

\section{The model}\label{sect:model}
\subsection{Two particle species with holes - a motivation}\label{sect:three_state}

In order to mimic bidirectional transport on a track we first introduce plus- and minus-directed particles on a one-dimensional lattice of $ L $ sites, each of which is either occupied by a plus (``$+$'') particle, occupied by a minus (``$-$'') particle, or empty. Empty sides are denoted as holes if the particle density $\rho<1$. Particles are identical up to their direction.
We have three types of stochastic, microscopic events between two neighboring sites, i.e.
\begin{eqnarray}
 + 0	 & \Rightarrow	& \ 0 + , \label{eq:plus0} \\
 0 - 		& \Rightarrow	& - 0 , \label{eq:0minus} \\ 
 + -		& \Rightarrow & - + . \label{eq:plusminus}
\end{eqnarray}
This two-species TASEP has two conserved quantities, i.e. the numbers $ N_{ \pm } $ of plus and minus particles,
under periodic boundary conditions.

 \begin{figure} [t!]
 	\centering
 	\includegraphics[width=1 \textwidth]{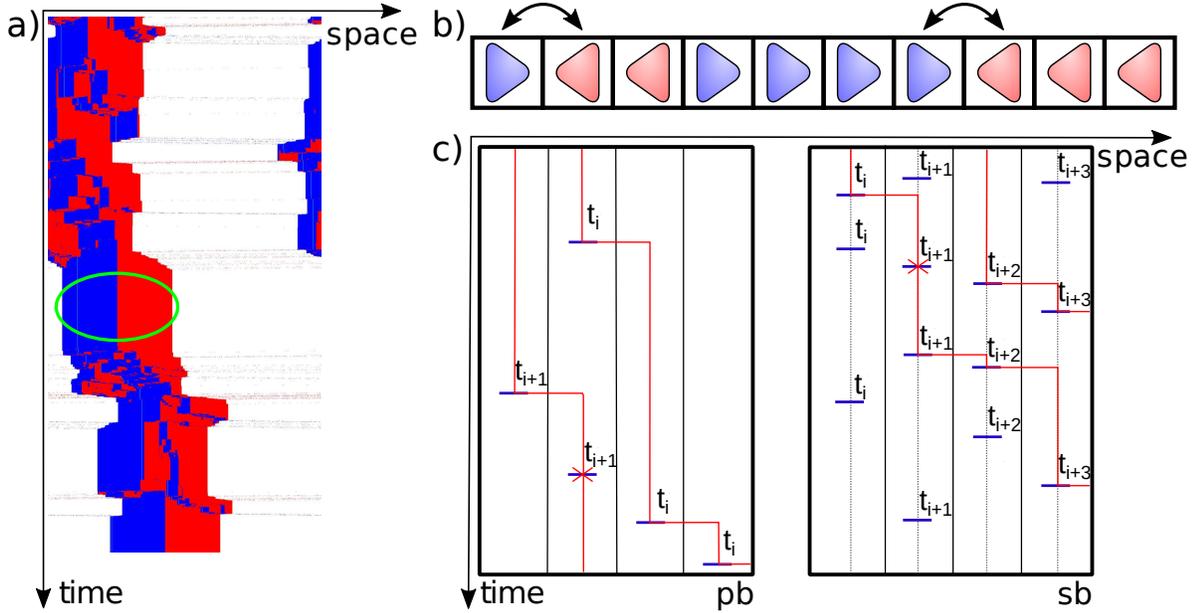}
 	\caption{\textbf{(a)} Kymograph section of a simulation of our two-species TASEP with holes, color-coded as plus particles in blue, minus particles in red and holes in white. 
 	Hopping times (\ref{eq:plus0}) and (\ref{eq:0minus}) are distributed exponentially, exchange times (\ref{eq:plusminus}) by a power law. The system size is $L=1000$ with $N_+=N_-=200$, the shown time interval is approximately 15000 time units. 
 	In this paper, we focus on the cluster, indicated by a green ellipse in the plot, leading to the model without holes defined in chapter \ref{sect:three_state}. 
 	\textbf{(b)} Scheme of the two-species model without holes. Triangles mimic particles, the tip indicates the direction. Exchange processes are only allowed for the configuration $(+-)$ (\ref{eq:plusminus}).
 	\textbf{(c)} Comparison of the two update schemes particle-based (left) and site-based (right) regarding particle trajectories (red lines). Timelines (dashed lines) for hopping events (blue bars) are carried by particles, hence identical to the red trajectory line in the left scheme or are fixed to the sites (right). Hopping events can be rejected due to exclusion (red crosses).
 	} 
 	\label{fig:kymo}
 \end{figure}

In a crowded environment, the exchange process (\ref{eq:plusminus}) can be very different from free hopping events and turn out to be the major criterium for estimating the particle flux similar to bottlenecks in unidirectional exclusion processes \cite{Greulich2008}. In a first approach we therefore impose Poissonian stochasticity on the normal jumps (\ref{eq:plus0}) and (\ref{eq:0minus}),
but a heavy tailed non-Markovian waiting time distribution on exchange processes (\ref{eq:plusminus}) in this section. It has been observed that the particle flux is heavily influenced by particle condensates, both in a two-species Markovian TASEP \cite{Evans1995} and in a one-species non-Markovian TASEP \cite{Concannon2014}. We see a similar phenomenon in first results of our two-species non-Markovian model, (figure~\ref{fig:kymo} (a)). 
Macroscopic clusters block particle flow for a long time interval interrupted by short boosts of particles hopping in a free space outside the clusters. This blockage is the major inhibitor of particle flux so that the waiting time distribution for exchange processes mainly controls the transport property of our system. We therefore focus on the cluster region indicated by the green ellipse in figure \ref{fig:kymo} (a) leaving only the two states ``$+$'' and ``$-$'' for a site in the lattice in the following.


\subsection{Two-species model without holes}\label{sect:two_state}

By concentrating on clusters and therefore neglecting holes, we consider a system fully occupied by plus and minus particles without holes $ N_+ + N_- = L $, for which an exemplary configuration is given in figure \ref{fig:kymo} (b). 

We remark that usually the two-species TASEP with $ N_+ + N_- = L $ is considered as standard TASEP below. We will explain the difference to the standard one-species TASEP in detail.

We first consider the standard one-species TASEP, where each site $i$ takes states $ \eta_i = 1 $ (occupied by a particle) or $ \eta_i = 0 $ (empty).
In most of the cases of the TASEP, the exponential distribution $ p(t)=\alpha^{-1}  e^{-\alpha t} $ $(t>0)$ is used to generate waiting times between two consecutive stochastic events, hence the TASEP is usually a Markov process. The parameter $\alpha$ stands for the hopping rate and is independent of time or the current system state.
A common way to assign waiting times to the Markovian TASEP is to use Gillespies \textit{direct method} or \textit{first reaction method} \cite{Gillespie1976}. In the latter one, times are generated from the exponential distribution for every possible hopping transition but only the smallest one determines the process which is executed and the other times are not used in the update mechanism anymore. In order to increase computational efficiency, a modified waiting time algorithm called \textit{next reaction method} \cite{Gibson2000} is storing the assigned times for every process and realizing them successively if the transition is allowed.

Using the next reaction method, we distinguish two ways to assign waiting times to the Markovian TASEP, illustrated in figure~\ref{fig:kymo} (c). The first one is a \textit{site-based} update. We generate and store a series of times $ \{ t_i^1 ,t_i^2 , \dots \} $ for each \textit{bond} between sites $ i $ and $ i+1 $. The difference $ t_i^{k+1} - t_i^k $ with $ t_i^0 = 0 $ must obey the exponential distribution. At time $t=t_i^k$, one checks whether the local configuration is appropriate for hopping, i.e. $\eta_i=1 $, $\eta_{i+1}=0 $. If this is the case, we move the particle from $i$ to $i+1$. If not, the move is rejected. The second one is a \textit{particle-based} update. We give a series of times $ \{ t_i^1 ,t_i^2 , \dots \} $ to each particle labeled by $ i $. Again the difference $ t_i^{k+1} - t_i^k $ should follow an exponential distribution. The particle $ i$ \textit{attempts} to hop to its right neighbor site at time $ t= t_i^k $, but the jump is again allowed only when the target site is empty. These two update schemes eventually yield equivalent dynamics to the particles, as long as we use an exponential distribution.


However, for a probability density function (PDF) $p(t)$ which belongs to a power law 
\begin{eqnarray}
 p (t) =\cases{
 0 & $ 0< t<1 $ , \\
 (\gamma-1) t^{ -\gamma} & $t > 1$,} 
 \label{eq:density}
\end{eqnarray}
with an exponent $\gamma > 1$ this equivalence does not hold anymore. We remark that the algebraic distribution violates the Markov property so that we use the next reaction method which is not equivalent to Gillespie's methods anymore. This is similar to the method used in \cite{Concannon2014, Gorissen2012} to evolve the system in time. See \ref{app:update} for details of the algorithm to perform simulations of our model. 

We recall that there are plus and minus particles, but no holes. 
As shown in figure~\ref{fig:kymo} (b), only the local exchanges of plus and minus particles are effective stochastic events in this case. This means that particle hopping refers to position exchange between neighboring particles. Instead of hopping of a single particle to an empty site, the exchange of particles in general depends on the status of both particles. We need more detailed rules to define the exchange process. 

The generalization of the site-based update to our non-Markovian TASEP is straightforward. 
We generate the time series $ \{ t_i^1 ,t_i^2 , \dots \} $ for each bond, such that now the difference $ t_i^{k+1} - t_i^k $ obeys the algebraic distribution (\ref{eq:density}). If we find $\eta_i= + $ and $\eta_{i+1}=- $ at time $ t= t_i^k $, we simply exchange the positions, i.e. we get a new configuration with $\eta_i= - $ and $\eta_{i+1}= + $. 

On the other hand, the particle-based update becomes a little complicated, and we need to further divide it into sub-schemes. The simplest one is considered as follows. We only assign the time series to the $+$ particles. At time $ t=t^k_i$, the $+$ particle labeled by $ i $ wants to hop rightward. This is allowed only if there is a $-$ particle on the target site. In other words, $-$ particles are regarded as holes, and the systems is  completely identical to the one-species non-Markovian TASEP that was introduced by Concannon \textit{et al.} Thus, we name this rule particle-based-\textit{asymmetrical} update, because this case does not hold plus-minus (or particle-hole after the identification) symmetry for Non-Markovian processes. 

Now we wish to look for rules which do not break the plus-minus symmetry to define two-species bidirectional transport using the particle-based approach.
We assign a time-series to each plus and minus particle $ \{ t_j^1 ,t_j^2 , \dots \} $. At time $t=t^i_j$ the particle is activated if a ``$+ -$'' configuration is given. We now consider two different local update schemes for a plus particle $i$ with waiting time $t_i^k$ and a minus particle with waiting time $t_j^m$ which are located at neighboring sites $l$ (plus particle) and $l+1$ (minus particle). 

1) The \textit{minimum rule}: An exchange between a pair of ``$+ -$'' particles is executed if one of the two particles is active. This means that the minimum of the two waiting times determines the particle exchange. 

2) The \textit{maximum rule}: An exchange of the particles is executed if both particles are active, i.e. a particle exchange happens after max$\{ t_i^k, t_j^m \}$. In both cases, the particles become passive after exchanging positions.


Let us summarize the four types of the update rules in the following: 
\begin{eqnarray} 
 \left\{ \begin{array}{ll} 
 \textrm{site-based}  \\ 
 \textrm{particle-based} 
 \left\{ \begin{array}{ll}
 \textrm{asymmetric}\\ 
 \textrm{symmetric} 
 \left\{ \begin{array}{ll}
 \textrm{maximum}  \\
 \textrm{minimum}  
 \end{array} \right. 
 \end{array}\right. 
 \end{array}\right. 
 \label{tab:four_types}
\end{eqnarray}

\section{Results} \label{sect:results}

In this section, we show simulation results of our non-Markovian two-species TASEP,
with a completely filled system, i.e. $N_+ + N_-= L$.
We first discuss the site-based update from \cite{Khoromskaia2014} as an algorithm for bidirectional transport in section \ref{sect:site-based} and then the previously described three types of particle-based rules in section \ref{sect:particle-based}.
For each of them we evaluate dynamics by measuring the PDFs for the \textit{effective} waiting time,
i.e. the duration between two realized exchange processes of a particle. 
We discuss the effective waiting time distribution, in the following two situations; the system with only one plus particle and $L-1$ minus particles called \textit{single particle dynamics} (spd), and the equally-filled case where $N_+=N_-$ called \textit{many particle dynamics} (mpd). We discuss the spd as a reference in order to highlight the collective effects modifying the original PDF $p(t)$ to the PDF for effective waiting times of exchange processes. With many particle dynamics, we investigate the influence of exclusion on the effective PDF.
In case of mpd, we also investigate transport efficiency by measuring particle flows. We compare the flows for the three symmetrical model rules in section \ref{sec:results_compare}.

\subsection{Site-based model} \label{sect:site-based}

The first bidirectional update we deal with is the site-based model inspired by \cite{Khoromskaia2014}. We start with spd, i.e. the situation where there is only one plus particle, i.e. $ N_+ = 1$ and $ N_- = L-1 $, and we probe its motion in the environment of minus particles. We give analytical estimates for and measure by simulations the effective waiting time distribution for site-exchange events, and compare the tail exponents of this quantity.

\begin{figure}[t!]
	\centering
	\includegraphics[width=.95 \textwidth]{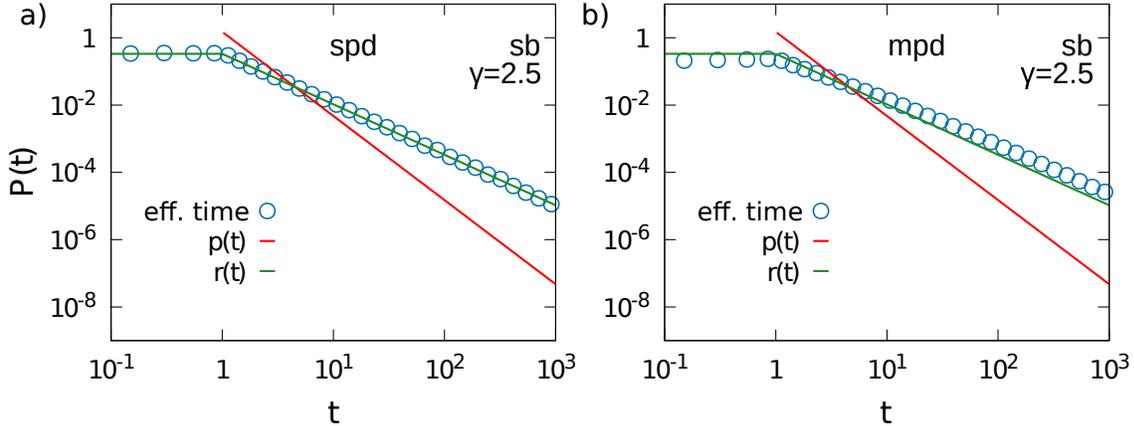}
	\caption{\textbf{Site-based rule: }PDFs of the effective exchange time for a system of $L=100$ for the site-based rule. (a) Single-particle dynamics, i.e. $N_+=1$ and $N_-=L-1$. (b) Many-particle dynamics, i.e.  $N_+=N_-=L/2$.}
	\label{fig:site-based}
\end{figure}

For the site-based update scheme, analytical estimates can be deduced from renewal theory as in \cite{Khoromskaia2014}. 
Let us assume that the plus particle arrives at a site $ i $, where an internal clock is already running since the last event on that site at time $t_i^{k}$, giving the clock an age $t-t_i^k$. The particle has to wait for the remaining time until $ t^{k+1}_i $ to execute the following step. This remaining time is called the \textit{residual} waiting time distributed by the PDF $r(t)$ which in general is different from the original PDF $ p(t)$. In the case where $ p(t) $ is given by equation (\ref{eq:density}), one finds the residual waiting time PDF in the form 
\begin{eqnarray}
\label{eq:residual_waiting_time}
r(t) =\cases{ 
 \textstyle \frac{\gamma-2}{\gamma-1} & $0<t<1$\\
 \textstyle \frac{\gamma-2}{\gamma-1} t^{1-\gamma} & $t\ge 1$ ,}
\end{eqnarray}
(see \ref{app:residual} for details). We note that $ r(t) > 0 $ even for $ 0 < t< 1 $, while the original $ p(t) $ is zero. More remarkably, the residual waiting time has a tail exponent shifted by 1, meaning that very large values have a higher statistical weight than the original PDF $ p(t)$.
In figure \ref{fig:site-based} (a), our simulation results for spd agree to the residual waiting time $r(t)$.

As a next step, we introduce $N_+=L/2$ plus particles, to check if the exclusion between plus particles further modifies the PDF. Results are shown in figure \ref{fig:site-based} (b) where the simulation results for exchange times also follow the asymptotics of $r(t)$ calculated for a single particle. This similar result is expected since the waiting times are renewed in both cases for every site in the lattice, no matter if a plus particle or a minus particle is occupying the site.

\begin{figure}[t!]
	\centering
	\includegraphics[width=0.35 \textwidth]{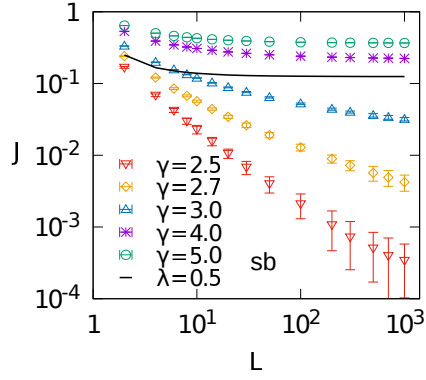}
	\caption{\textbf{Site-based rule:} Length dependency of the particle flux $J$ for different values of $\gamma$ in the PDF (\ref{eq:density}) and mpd. The black line serves as a comparison to the flux generated by an exponential distributed waiting times with exponent $\lambda$. }
	\label{fig:length_dependency_sb}
\end{figure}
\begin{figure}[b!]
	\centering
	\includegraphics[width=.95 \textwidth]{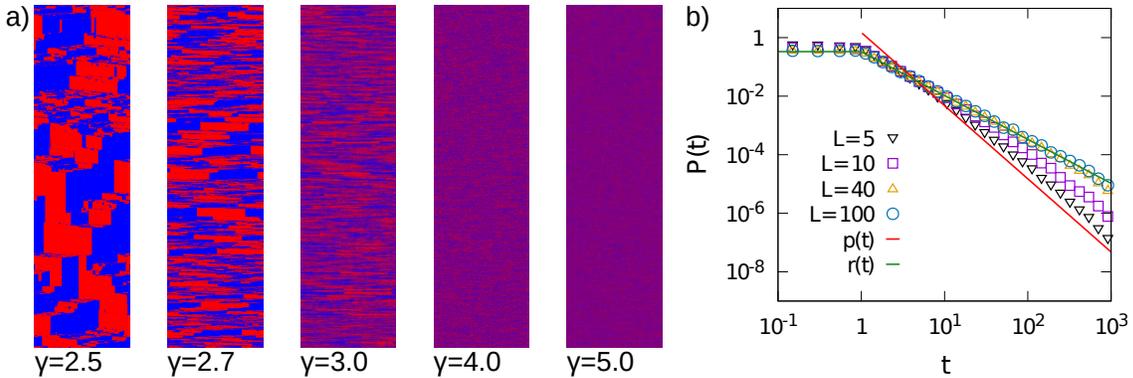}
	\caption{ (a) Kymographes for the site-based model with runtime: $8 \cdot 10^{10}$ system size  $L=1000$ and equal particle numbers $N_+=N_-$. For $\gamma<3$ it is computationally hard to achieve stationary states. (b) Finite size effects in PDFs for spd following the site-based model with an exponent $\gamma=2.5$.}
	\label{fig:length_dependency_new}
\end{figure}

After discussing the influence of the used update method and the particle interaction on the PDF of exchange times, we will continue by examining transport properties in a bidirectional, two-species system. As discussed of effective exchange time distributions, we will continue by examining how these effects are reflected in the transport properties in the bidirectional two-species TASEP with $N_+=N_-=L/2$.

A common way to measure the transport efficiency of the system is the particle flux $J$. In \cite{Concannon2014}, it was shown that for the unidirectional non-Markovian TASEP, system size has an impact on $J$ in the clustering phase. Since we find similar shifts of the PDF for the site-based update, we also expect a transition from a size-dependent to a size-independent regime with growing $\gamma$. We test this transition by plotting the particle flux versus the length of the system in figure \ref{fig:length_dependency_sb} for different exponents $\gamma$. Up to a finite-size effect, no significant dependency of $J$ is observed when using a $\gamma>3$ in the simulations. The flow converges for large $L$, so that no significant size-dependency is observed for $L>100$.

However, $J$ is decreasing with the system size for $\gamma<3$. At the same time, the error bars are larger and it is difficult to judge the limit for infinite system size from numerical results. It is, therefore, necessary to argue with additional information about a transition from length dependency to constant fluxes. From the residual waiting time, we know that the average effective waiting time diverges at $\gamma=3$. So we expect the flux to vanish below this critical value $\gamma_c=3$ for infinite systems similar as pointed out in \cite{Khoromskaia2014}.
Below $\gamma_c$, the transport efficiency is determined by the asymptotics of the effective waiting time distribution. This effect induces a strong size dependence of the flow and at the same time increases the relaxation times of the system.


In order to illustrate the computational complexity, we show kymographes of a system of length $L=1000$ which cover a time period of $8 \cdot 10^{10}$ time units for different exponents in figure \ref{fig:length_dependency_new} (a). Homogeneous spatiotemporal structures are obtained for $\gamma \geq 4$, while stable clusters emerge for $\gamma \leq 3$. In this regime, the lifetime of the clusters are comparable to the simulation time, which make it difficult to reach the stationary state of the system numerically. The slow relaxation of the system is caused by rare extreme values for the waiting times which develop in an aging phase inside a cluster and block other exchange events such as in \cite{Concannon2014}. 
We estimate the uncertainty of $J$ by using the partial time averages $J_n=\left(\sum_{t=t_n}^{t_n+\Delta} J_t \right)/ \left(\sum_{t=t_n}^{t_n+\Delta} t\right) $ where $\Delta$ is the complete time of measurement divided by the number of partial blocks and $t_n=n\Delta$.


For the site-based model, we expect that the residual waiting time PDF (\ref{eq:residual_waiting_time}) describes exactly the effective exchange time distribution. The results of our simulations, however, slightly differ from this prediction. We expect that the small deviation can be attributed to a finite-size effect, since local waiting times may exceed the typical time a particle needs for a complete tour in a finite periodic system. This finite-size effect is observable in figure \ref{fig:length_dependency_new} (b) where tails differ from $r(t)$ for small system sizes.

\subsection{particle-based models} \label{sect:particle-based}
\subsubsection{Asymmetrical rule} \label{sect:asym}

As a reference system for particle-based updates, we start with the asymmetrical rule, which is identical to the model of Concannon \textit{et al.} \cite{Concannon2014}. In the paper, it was pointed out that interaction via exclusion leads to a shift of one for exponents in the hopping time PDF for a unidirectional many particle system. Again, we will show PDFs for single-particle and many-particle dynamics. 

\begin{figure}[b!]
	\centering
	\includegraphics[width=.95 \textwidth]{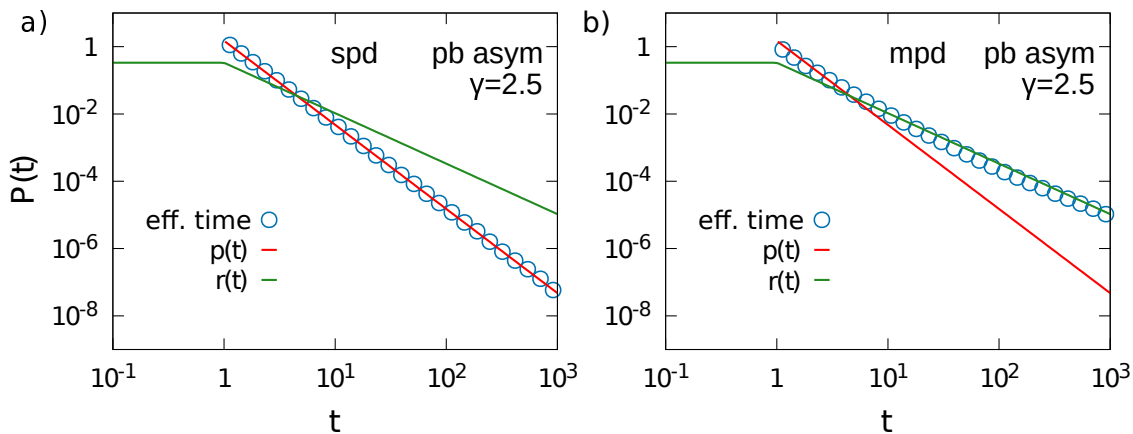}
	\caption{\textbf{Particle-based asymmetrical rule:} PDFs of the effective exchange time for a system of $L=100$ for the asymmetrical particle-based rule. (a) Single-particle dynamics, i.e. $N_+=1$ and $N_-=L-1$. (b) Many-particle dynamics, i.e.  $N_+=N_-=L/2$.}
	\label{fig:particle-based_asym}
\end{figure}

The spd under the particle-based, asymmetric update rule is much easier than for the site-based rule. The time-series $\{ t_1^1,t_1^2, \dots \}$ contains the times for which the plus particle moves, since it is never blocked by other plus particles. In other words, the effective waiting time distribution is $ p(t) $ itself, and we can regard the plus particle simply as a non-Markovian random walker while minus particles serve as passive holes. The motion of the particle is completely determined by $\{ t_1^1,t_1^2, \dots \}$. Our simulation results for this scenario are shown in figure~\ref{fig:particle-based_asym}.

Now turning to the mpd realization of the asymmetric update, we get back the scenario discussed in \cite{Concannon2014}, since minus particles are passive and correspond to holes of the one-species TASEP. For completeness, the results are shown in figure \ref{fig:particle-based_asym} (b). 
As expected, we observe that for small times near $t=1$, the simulation data points follow the original density $p(t)$ but then the exponent of PDF changes to $\gamma-1$ for larger times as predicted in \cite{Concannon2014}. 
The particle flux and its dependency on the system size for $\gamma \leq \gamma_c$ has already been discussed in \cite{Concannon2014}, hence we continue with the symmetric rules for particle-based updates.

\subsubsection{Maximum rule} \label{sect:max}

Let us turn to the first particle-based-symmetrical update, the maximum rule, which was introduced in section \ref{sect:model}. This rule does not break the symmetry between plus and minus particles because the exchange process is triggered by both particles (with index $i$ and $j$) that have to be activated for an exchange process first. Hence, one has to choose the maximum from the two next event times in the time-series $ { t_i^k, t_j^\ell } $, in order to determine when the plus particle actually exchanges with its neighbor. We estimate PDFs for the symmetrical maximum update by calculating the density of the maximum of two random variables $X$ and $Y$ with density $p_1(t)$ and $p_2(t)$. 

Because all particles follow to their own time series $ \{ t_i^1 ,t_i^2 , \dots \}$ and $ \{ t_j^1 ,t_j^2 , \dots \}$, different situations appear for neighboring pairs of plus and minus particles. There is always one particle inducing the exchange process, e.g. the one with the later time for the maximum update rule. However, the process can be induced by a plus particle or a minus particle.

Let us assume that a plus particle induced an exchange $k$ and then becoming involved in a new exchange process $k+1$ with a new partner. Here, the plus particle's clock has no age immediately after the last executed exchange process. In contrast, the new neighbor minus particle already is located on its position for some time meaning a clock with a non zero age. Hence, the plus particle follows the density $p_{plus}(t)=p(t)$ but we assume that the minus particle's residual waiting time is rather described by $p_{{minus}}(t)=r(t)$ because it is standing in the queue of other minus particles. In the single particle case, the maximum of two random times  is taken from a time $X$ that follows $p(t)$ and a time $Y$ that follows $r(t)$ as $max(X_{p(t)},Y_{r(t)})$ which we call a \textit{mixed scenario} in the following.

However, for symmetric rules, minus particles also can introduce an exchange process. Let us follow a plus particle again, but this time the exchange $k$ was induced by its former neighboring particle. The plus particle has a new exchange partner for the process $k+1$ again, but this time also a non zero age, just as its new neighbor minus particle which was not involved in the former process $k$. In this case, we assume both residual waiting times are distributed by $p_{plus}(t)=p_{minus}(t)=r(t)$. The maximum of two random times $X$ and $Y$ is now chosen as $max(X_{r(t)},Y_{r(t)})$, called the \textit{pure scenario}.

First, we give the cumulative distribution function (CFD) for both $p(t)$ and $r(t)$:
\begin{eqnarray}
\label{eq:CDFp}
P(t) =\cases{ 
	\textstyle 0& $0<t<1$\\
	\textstyle 1- t^{1-\gamma} & $t\ge 1$,}
\end{eqnarray}
\begin{eqnarray}
\label{eq:eq:CDFr}
R(t) =\cases{ 
	\textstyle \frac{\gamma-2}{\gamma-1}t& $0<t<1$\\
	\textstyle 1- \frac{1}{\gamma-1}t^{2-\gamma} & $t\ge 1$.}
\end{eqnarray}
With these, we can now calculate the density of the maximum $X$ and $Y$, i.e.
\begin{eqnarray}
f_{max}^{mix}(t) &= \textstyle \frac{d}{dt}\left[ P(t)R(t) \right] \\
&=	\cases{ 
	\textstyle 0 & $0<t<1$\\
	\textstyle (\gamma-1)t^{-\gamma} + \frac{\gamma-2}{\gamma-1} t^{1-\gamma} + \frac{3-2\gamma}{\gamma-1} t^{2(1-\gamma)}& $t\ge 1$,}
\end{eqnarray}
\begin{eqnarray}
f_{max}^{pure}(t) &=  \textstyle \frac{d}{dt}\left[ R(t)R(t) \right]  \\
&=	\cases{ 
	\textstyle 2\left(\frac{\gamma-2}{\gamma-1}\right)^2 t & $0<t<1$\\
	\textstyle 2 \frac{\gamma-2}{\gamma-1} t^{1-\gamma}  - 2 \frac{\gamma-2}{\left(\gamma-1\right)^2} t^{3-2\gamma}  & $t\ge 1$.}
\end{eqnarray}
\begin{figure}[b!]
	\centering
	\includegraphics[width=.95 \textwidth]{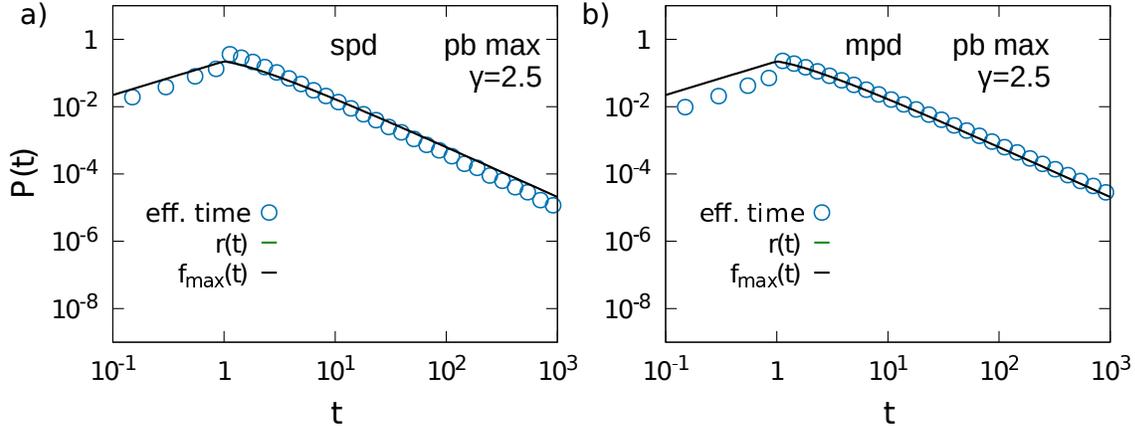}
	\caption{\textbf{Particle-based maximum rule:} PDFs of the effective exchange time for a system of $L=100$ for the particle-based maximum rule. (a) Single-particle dynamics, i.e. $N_+=1$ and $N_-=L-1$. (b) Many-particle dynamics, i.e.  $N_+=N_-=L/2$.}
	\label{fig:particle-based_max}
\end{figure}
Both, the pure and mixed scenarios have the leading exponent $1-\gamma$ which is equal to the exponent of $r(t)$.
In the pure scenario, we get an estimate for effective waiting times smaller than 1. This expression is expected to overestimate the weight of the waiting times because also the mixed scenario is contributing to the exchange processes, always with times larger than one. However, we will use the pure scenario as an estimate for the spd effective waiting time density for the maximum rule in the following, 
\begin{eqnarray}
\label{eq:max_waiting_time}
f_{max}(t)=f_{max}^{pure}(t) =\cases{ 
	\textstyle 2\left(\frac{\gamma-2}{\gamma-1}\right)^2 t & $0<t<1$\\
	\textstyle 2 \frac{\gamma-2}{\gamma-1} t^{1-\gamma}  - 2 \frac{\gamma-2}{\left(\gamma-1\right)^2} t^{3-2\gamma}  & $t\ge 1$.}
\end{eqnarray}

We compare these estimates with simulation results in figure \ref{fig:particle-based_max} (a) for spd. Here, the tail behavior of our simulation results are in good agreement with $f_{max}(t)$. Also, short time behavior is well approximated when relating the sharp increase at $t\approx1$ to the influence of the mixed scenario which estimates a zero probability for $t<1$ and a higher weight at $t>1$ for the used $\gamma=2.5$.

We now compare $f_{max}(t)$ to simulation results of the mpd in figure \ref{fig:particle-based_max} (b). Again we see that the tail behavior is well described by $f_{max}(t)$ and $r(t)$. The short time behavior is still close to the estimate but the tail has a higher statistical weight comparing to spd. 

The shift in the exponent that is seen when comparing the original waiting time PDF and the effective exchange time PDF is also consistent with the results for the flux in the maximum rule which is shown in figure \ref{fig:length_dependency_minmax} (b). We observe no significant changes in the flux for system sizes larger than $L=100$ if $\gamma>\gamma_c$ but a flux that vanishes with $L$ for $\gamma<\gamma_c$, similar to the results found for the site-based model.

\subsubsection{Minimum rule} \label{sect:min}

The second particle-based-symmetrical update is the minimum rule, also introduced in section \ref{sect:model}. Also, this rule does not break the symmetry between plus and minus particles. The exchange process is triggered by the first particle that is activated for an exchange process at the minimum time of the two next event times in each particle's time-series $ { t_i^k, t_j^\ell } $. We calculate an estimates for PDFs from the minimum of two random variables $X$ and $Y$ with density $p_1(t)$ and $p_2(t)$. 

As in the maximum model, we have plus induced and minus induced exchanges. In our analytical estimate we use again the assumption that the particle with an aged residual waiting time is distributed by $r(t)$ so that we have to calculate $min(X_{p(t)},Y_{r(t)})$ for the mixed scenario and $min(X_{r(t)},Y_{r(t)})$ for the pure scenario.

\begin{figure}[b!]
	\centering
	\includegraphics[width=.95 \textwidth]{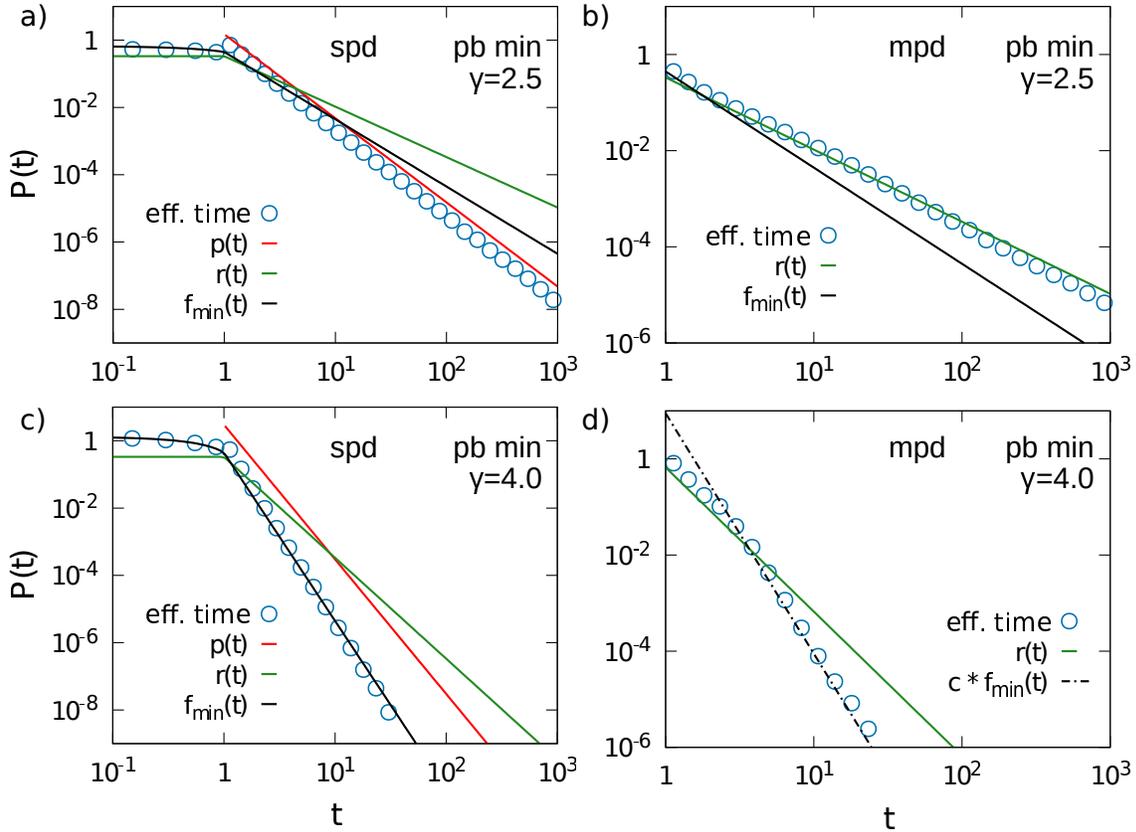}
	\caption{\textbf{Particle-based minimum rule:} PDFs of the effective exchange time for a system of $L=100$ for the particle-based minimum rule. (a) Single-particle dynamics, i.e. $N_+=1$ and $N_-=L-1$ for exponent $\gamma=2.5$. (b) Many-particle dynamics, i.e.  $N_+=N_-=L/2$ for exponent $\gamma=2.5$. (c) Single-particle dynamics for exponent $\gamma=4.0$. (d) Many-particle dynamics for exponent $\gamma=4.0$. The dashed line shows the tail behavior of $f_{min}(t)$ close to the simulation data.}
	\label{fig:particle-based_min}
\end{figure}

We again use the CFDs of equations (\ref{eq:CDFp}) and (\ref{eq:eq:CDFr}) to calculate the minimum density in both scenarios:
\begin{eqnarray}
f_{min}^{mix}(t) &= \textstyle p(t)\left[ 1-R(t) \right] + r(t)\left[ 1-P(t) \right]\\
&=	\cases{ 
	\textstyle \frac{\gamma-2}{\gamma-1} & $0<t<1$\\
	\textstyle \frac{2\gamma-3}{\gamma-1} t^{2-2\gamma} & $t\ge 1$,}
\end{eqnarray}
\begin{eqnarray}
f_{min}^{pure}(t) &=  \textstyle 2r(t)\left[ 1-R(t) \right] \\
&=	\cases{ 
	\textstyle 2\frac{\gamma-2}{\gamma-1}\left( 1- \frac{\gamma-2}{\gamma-1}t \right)& $0<t<1$\\
	\textstyle 2 \frac{\gamma-2}{\left(\gamma-1\right)^2} t^{3-2\gamma} & $t\ge 1$.}
\end{eqnarray}
We realize that the largest tail is $3-2\gamma$ which we get for the pure scenario. 
Since both scenarios can be observed in the exclusion process, each of them contributes to the effective exchange time but the tail behavior is determined by the slower process that can block transport completely. We therefore set
\begin{eqnarray}
\label{eq:min_waiting_time}
f_{min}(t) =f_{min}^{pure}(t)=
\cases{ 
	\textstyle 2\frac{\gamma-2}{\gamma-1}\left( 1- \frac{\gamma-2}{\gamma-1}t \right)& $0<t<1$\\
\textstyle 2 \frac{\gamma-2}{\left(\gamma-1\right)^2} t^{3-2\gamma} & $t\ge 1$.}
\end{eqnarray}

Note that $f_{min}(t)$ leads to a increase in transport efficiency for $\gamma>3$ but the exponent of $f_{min}(t)$ becomes larger than $-\gamma$ if $\gamma>3$. This counter-intuitive result follows from the assumption that both waiting times of interfacing particles are distributed by $r(t)$ instead of $p(t)$. However, the prediction would mean a slower exchange than in the asymmetric particle-based rule where minus particles are completely passive. We will see that our estimates actually describes the simulation results for spd only for $\gamma>3$ in figure \ref{fig:particle-based_min} (c) but not in panel (a) where $\gamma=2.5$. Here, the tail behavior is well represented by $p(t)$.

In order to understand the origin of this difference in tail exponents, we measure the residual waiting time carried by the plus particle, which we call $\tilde{p}(t)$. We find that in the tail $\tilde{p}(t) \approx r(t)$ for $\gamma>3$ but not for $\gamma<3$ where the exponent is not exceeding values of $\gamma=-2$. Calculating the minimum with such an exponent from $\tilde{p}$ would lead to an $f_{min}(t)$ with exponents $\leq -\gamma$, i.e. $p(t)$ serves as a upper limit (see \ref{app:minmax} for details).

\begin{figure}[t!]
	\centering
	\includegraphics[width=0.7 \textwidth]{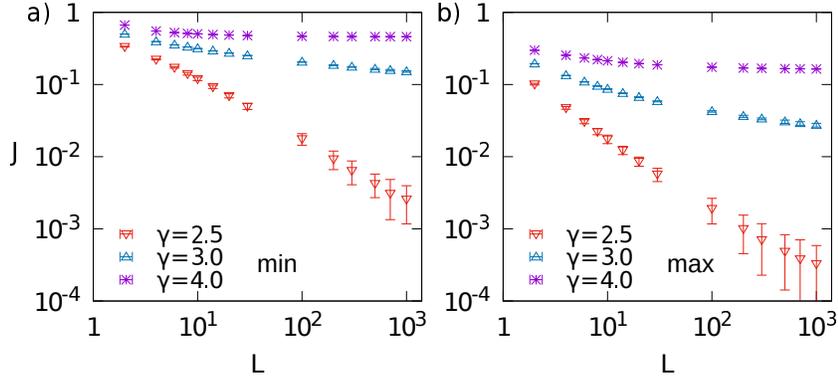}
	\caption{\textbf{(a) Minimum and (b) maximum rule:} Length dependency of the particle flux $J$ for different values of the exponent $\gamma$ in the PDF (\ref{eq:density}) and mpd. }
	\label{fig:length_dependency_minmax}
\end{figure}

We can understand this deviation by realizing that the plus induced dynamics is getting more important for if $\gamma$ is below the critical value $\gamma_c$. The influence of the tail in the residual waiting time is important for inducing events by minus particles that did stand in the queue for a long time. In contrast, the plus particle is more often responsible for inducing the events and consequently determines the effective exchange time. We show that the ratio of plus induced events is growing in this regime in \ref{app:minmax}. The time average in the calculation of the residual waiting time in \ref{app:residual} is not valid due to temporal correlations in this scenario.

For mpd, similar behavior is observed. The estimate $f_{min}$ is well suited to the simulation result if $\gamma>3$, which is shown in figure \ref{fig:particle-based_min} (d). In results for small exponents $\gamma<3$ shown in figure \ref{fig:particle-based_min} (b), the analytical estimate again does not describe the simulation results. Instead, the tail is determined by the residual waiting time $r(t)$.
The relevance of the residual waiting time is caused by the dominance of long exchange times in long queues. Furthermore, passively exchanged particles keep their event time after an exchange process which leads to long range correlations of particles exchange times.
 
Similar to the results of the site-based model and the maximum model, the flux in the minimum rule does not show significant changes with $L$ in the fast decaying regime above $\gamma_c$, which is shown in figure \ref{fig:length_dependency_minmax} (a). For $\gamma<\gamma_c$, again a dependency on the system length in the data supports the qualitative difference between the regimes found in results from effective exchange times above. However, the flux clearly is higher in the minimum rule than in the maximum rule (panel b)). We want to further study the difference of the applied model rules on the particle flux in the next section.

\subsection{Transport efficiency of symmetric model rules}
\label{sec:results_compare}

\begin{figure}[t!]
	\centering
	\includegraphics[width=0.9 \textwidth]{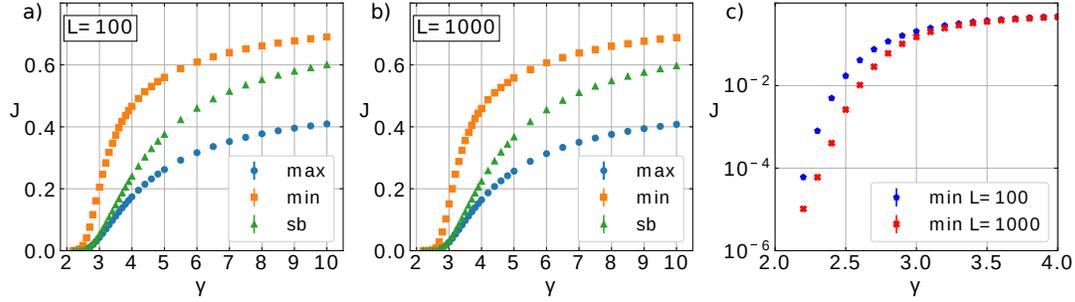}
	\caption{Particle flux depending on the exponent $\gamma$ for the three different update rules (site-based green, minimum orange, maximum blue). Error bars are drawn from sem values of 10 realizations. (a) The system size is $L=100$, (b) $L=1000$. (c) Section for $\gamma\leq 4$ in a logarithmic scale for $J$ for $L=100$ (blue) and $L=1000$ (red). }
	\label{fig:flux_comparison}
\end{figure}

The exponents for effective exchange times found for the different model rules are summarized in table \ref{tab:four_types_shift}. We now compare the flux which is generated by the three symmetrical updates in figure \ref{fig:flux_comparison}. Even though the asymptotic behavior of the three rules are similar (except for $\gamma>\gamma_c$ in the minimum model), the short exchange times influence the value of the flux. This leads to significant differences between the site-based model, the particle-based minimum model and the particle-based maximum model for $\gamma>\gamma_c$. As expected, the maximum rule is really slower than the site-based model and the minimum rule can enhance the transport. For $\gamma<\gamma_c$ however, the flux generated in the maximum rule is close to the flux by the site-based rule. In the minimum rule, we still measure higher fluxes, both for the system size of $L=100$ (a) and $L=1000$ (b). Note that we observe finite-size effects in these results which is shown in panel c) where the data points deviate for the different system sizes. By the analysis of tail exponents, we expect $J=0$ for the infinite system in the stationary state such as in the other models.

\begin{table}[b!]
	\caption{The resulting exponent seen in the effective waiting time PDFs in the different update rules for single-particle dynamics and many-particles dynamics.}
	\begin{center}					
		\begin{tabular}{|c|c|c|}	\hline 
			& single-particle dynamics & many-particle dynamics \\ 
			\hline 
			site-based & $1-\gamma$ & $1-\gamma$ \\ 
			\hline 
			particle-based asymmetrical  & $-\gamma$ & $1-\gamma$ \\ 
			\hline 
			particle-based maximum & $1-\gamma$ & $1-\gamma$  \\
			\hline
			particle-based minimum & 
			\begin{tabular}{c|c} 
				$-\gamma$   & $3-2\gamma$	\\ 	
				for $\gamma<3$  &  for $\gamma>3$
			\end{tabular}  
			& 	
			\begin{tabular}{c|c} 
				$1 -\gamma$   & $3-2\gamma$	\\ 	
				for $\gamma<3$  &  for $\gamma>3$
			\end{tabular}   \\ 
			\hline 
			
		\end{tabular} 	
	\end{center}	\label{tab:four_types_shift}  
\end{table}

\section{Conclusion }\label{sect:conclusion}

In our contribution, we analyzed different bidirectional variants of the TASEP with non-Markovian exchange dynamics. These models are relevant for one-dimensional transport problems in crowded environments, where the high density of particle clusters leads to small effective exchange rates of particle positions. 
A possible realization of the bidirectional transport model would include two oppositely moving particle species and holes. 
A model of this kind would combine a Markovian particle-dynamics, which would be applied when the particles move toward an empty site and a non-Markovian 
particle-dynamics, which governs the particle-exchange. Simulation results show strong condensation of the particles, which implies that the bidirectional transport 
capacity is determined by the efficiency of the exchange processes rather than by the time spent in the low density area. 
Therefore, we restricted our analysis to symmetric and fully-filled systems. This choice reduces considerably the corrections to scaling for small system sizes.  

Modeling bidirectional transport of active particles with lattice gases allows assigning the exchange times to the particles as well as to the lattice. In the latter case, 
we can map the problem to the uni-directional process, since pairs of oppositely moving particles behave as particles and holes in the uni-directional case. This is even true for spd which correspond to a uni-directional system with a single hole where the dynamics of the particle is governed by the residual waiting time.
Significant differences to the uni-directional case exist if the exchange times are assigned to the particles. This can be realized in a symmetric or in an asymmetric way,
wherein the latter case the reaction times are assigned to only one-particle species. The asymmetric case implies that one particle species can be assumed to be passive which is why we find the asymptotics of the uni-directional non-Markovian TASEP for the single-particle as well as for the many-particle dynamics.

Assigning exchange times symmetrically to both particle species implies that two exchange times are given for a pair of oppositely moving particles. Therefore, one has to define an additional selection rule. In this work we have chosen two extreme cases that preserve the symmetry between the two types of particles, i.e. either the minimum or the maximum of the two waiting-times will be selected. 
In case of the maximum rule, residual waiting time and $f_{max}(t)$ show the same asymptotics. Therefore the maximum rule modifies indeed the effective exchange time distribution but to the exponent compared to the residual waiting time of the site-based model. Significant differences exist only in the $\gamma > \gamma_c$ regime for the minimum rule. In this case,  we find that the asymptotics is governed by the distribution of independent residual times where the asymptotics of effective exchange time distribution is given by $3-2\gamma$. For $\gamma < \gamma_c$ this is not the leading contribution. Here, the asymptotic behavior is in accordance with the asymmetric particle based model which is given by the residual waiting time for mpd. This effect is the result of passively moving particles which keep the assigned exchange time. For small values of $\gamma$ the dynamic is, as for the other cases, governed by pairs of particles with long residual waiting times.  
 
Our results underline the universality of the findings which have been discussed for the uni-directional non-Markovian TASEP \cite{Concannon2014, Khoromskaia2014}. In this class of models, many particle effects generically lead to a dominant contribution of the residual waiting time for $\gamma < \gamma_c$.  Here, the configurations are characterized by large particle clusters and a size dependent flow of particles. The transport capacity of large systems in this parameter regime is extremely low compared to their Markovian counterparts. For $\gamma > \gamma_c$ however, we observe homogeneous particle configurations and size-independent values of the flow which differ for the different implementations of the dynamics. 

Our findings can be relevant for bidirectional flows under strong confinement as for example in narrow escape problems in pedestrian dynamics \cite{Schadschneider2010} or intracellular transport in axons and dendrites \cite{Shemesh2010} where the effective exchange dynamics can be non-Markovian.

\ack

	\noindent This work was funded by the Deutsche Forschungsgemeinschaft (DFG) through Collaborative Research Center SFB 1027 (Project A8).\\
\newpage
\appendix
\section{Calculation of residual waiting times}\label{app:residual}

The waiting time PDF for a single particle site exchange in the site-based model is calculated by using renewal theory following \cite{Gallager2013}. In particular, we will determine the residual waiting time until the next exchange event occurs if two particles are in the local $(+-)$ configuration. We start with a renewal process for renewal waiting times $X_n$ distributed by Eq.\,(\ref{eq:density}), which are given to a site in the lattice.  The $N$-th renewal of the waiting time on this site occurs at time 
\begin{eqnarray}
S_{N} = \sum_{n=1}^{N} X_n,
\end{eqnarray}
i.e. we can count the number of passed renewal events $N(t)$ at each time $t$. 

\begin{figure}[b!]
	\centering
	\includegraphics[width=.9 \textwidth]{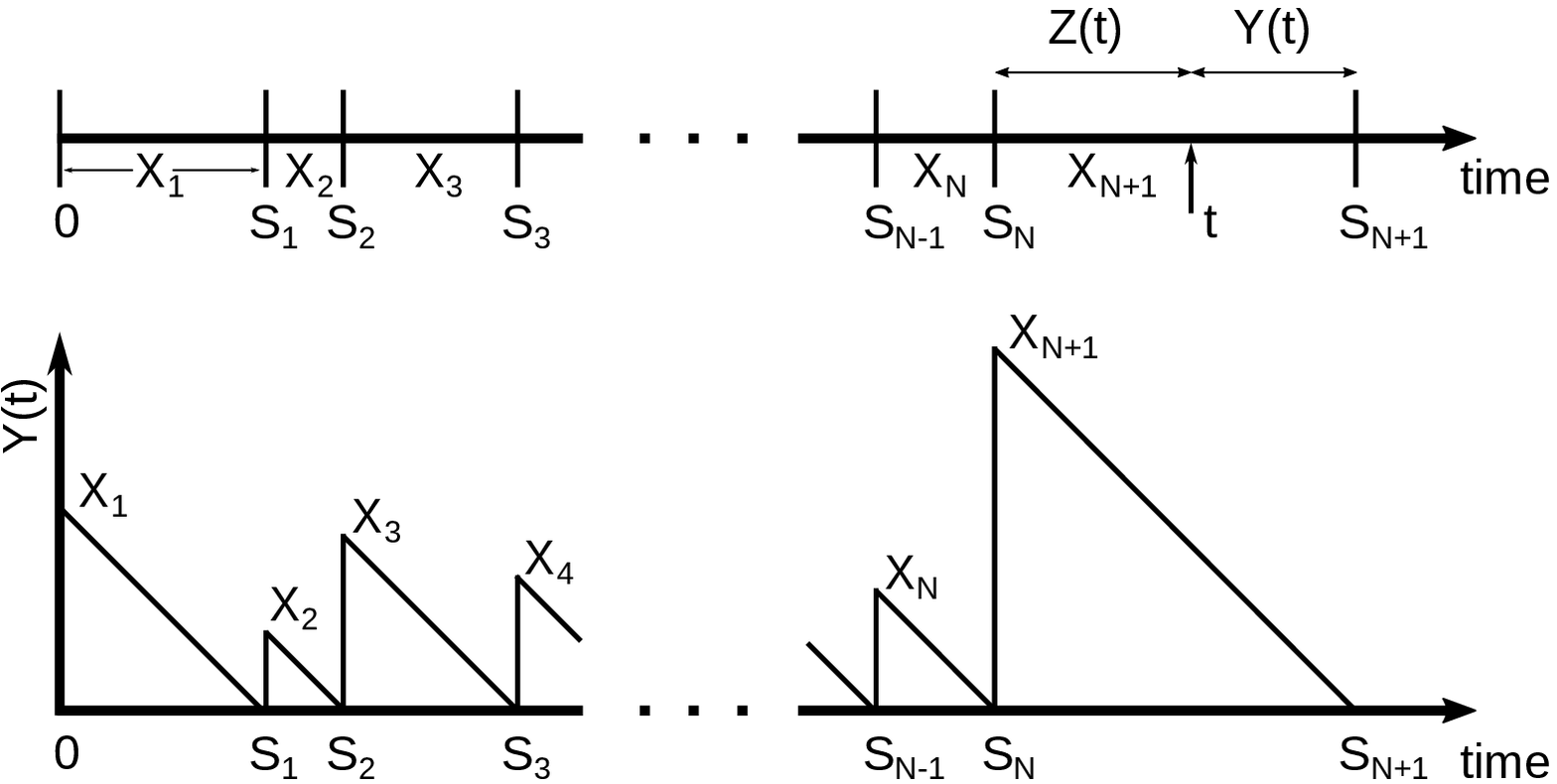}
	\caption{\textbf{Top:} The renewal process is determined by the time series $X_n,$ $n\in \mathbb{N}$, build from the algebraic waiting time PDF in Eq.\,(\ref{eq:density}). Summing up these waiting times $S_1, S_2, ...\, S_N$ gives the time for the next event at $S_{N+1}$. For a time $S_n \leq t \leq S_{N+1}$, the current process has the age $Z$ and the residual life $Y$. \textbf{Bottom:} The residual life $Y(t)$ is a step wise function of time, decaying from $X_n$ to $0$ during the time in the interval between $S_{n-1}$ and $S_n$. }
	\label{fig:residual_scheme}
\end{figure}

For a time $t>S_{N(t)}$, the waiting times of this site are called the duration of renewal time intervals $\tilde{X}(t) = X_{N(t)+1} = S_{N(t)+1}-S_{N(t)}$ (see figure \ref{fig:residual_scheme}). The renewal process also has an age $Z(t) = t - S_{N(t)}$ as well as a residual life (residual waiting time) $Y(t) = S_{N(t)+1}-t$ until the next renewal event takes place at time $S_{N(t)+1}$. The residual waiting time is therefore also written as $Y(t)=\tilde{X}(t)-Z(t)$.

We now calculate the time averaged CDF of the residual waiting time $Y(t)$, i.e.  $F_{Y}(y) = {\rm Pr} \{ Y(t) \leq y \} $ that gives the fraction of time that the residual waiting time is smaller than a given $y$. We can invent an indicator reward function $R(t)$ to determine if the residual waiting time is actually smaller or not, i.e. 
\begin{eqnarray}
R(t) = R(Z(t), \tilde{X}(t))= \cases{1 & for $\tilde{X}(t)-Z(t) \leq y$ \\
	0 & otherwise. \\}
\end{eqnarray}
The following form for the CDF of $Y(t)$ can then be found by using the time average over the indicator function 
\begin{eqnarray}
\label{eq:residual}
F_{Y}(y) = \lim_{t\rightarrow \infty}\frac{1}{t}\int_0^t R(\tau) \rmd \tau = \frac{1}{\overline{X}} \int_{x=0}^{x=y} {\rm Pr} \{ X>x \} \rmd x,
\end{eqnarray}
where $\overline{X}=\frac{\gamma-1}{\gamma-2}$ denotes the mean value of the renewal event duration for the PDF in Eq.\,(\ref{eq:density}).

We now use this framework to determine the residual waiting time for the renewal process with algebraic waiting time PDF Eq.\, (\ref{eq:density}), i.e. 
\begin{eqnarray*}
	f_X(x) =\cases{
		0 & $ 0< x<1 $ , \\
		(\gamma-1) x^{ -\gamma} & $x > 1$.} 
\end{eqnarray*}
The CDF of the renewal time intervals is 
\begin{eqnarray}
F_X(x) = \cases{0 & $x < 1$ \\
	1 - x^{1-\gamma} & $ x \ge 1$.\\ }
\end{eqnarray}
We use equation \ref{eq:residual} to determine the CDF
\begin{eqnarray}
F_Y(y) &= \frac{1}{\overline{X}} \int_{0}^{y} \left( 1 - (1-x^{1-\gamma})\Theta(x-1) \right) \rmd x \\
&=\cases{\frac{\gamma-2}{\gamma-1}~ y & $y<1$\\
	\frac{\gamma-2}{\gamma-1} \left( \frac{\gamma-1}{\gamma-2} + \frac{1}{2-\gamma}y^{2-\gamma} \right) & $y\ge 1$, }
\end{eqnarray}
and finally arrive at the result
\begin{eqnarray}
Y(y) =\cases{\frac{\gamma-2}{\gamma-1} & $y<1$\\
	\frac{\gamma-2}{\gamma-1} ~ y^{1-\gamma} & $y\ge 1$}
\end{eqnarray}
for the PDF of the residual waiting time $Y(y)$. 

\newpage
\section{Additional measurements for the minimum rule in single-particle dynamics}\label{app:minmax}
In this appendix, we further examine results in the single-particle dynamics, minimum update rule. As we have seen in figure \ref{fig:particle-based_min}, the effective waiting time distribution follows the estimation $f_{min}^{pure}$ only for exponents $\gamma>3$. 

\begin{figure}[b!]
	\centering
	\includegraphics[width=.95 \textwidth]{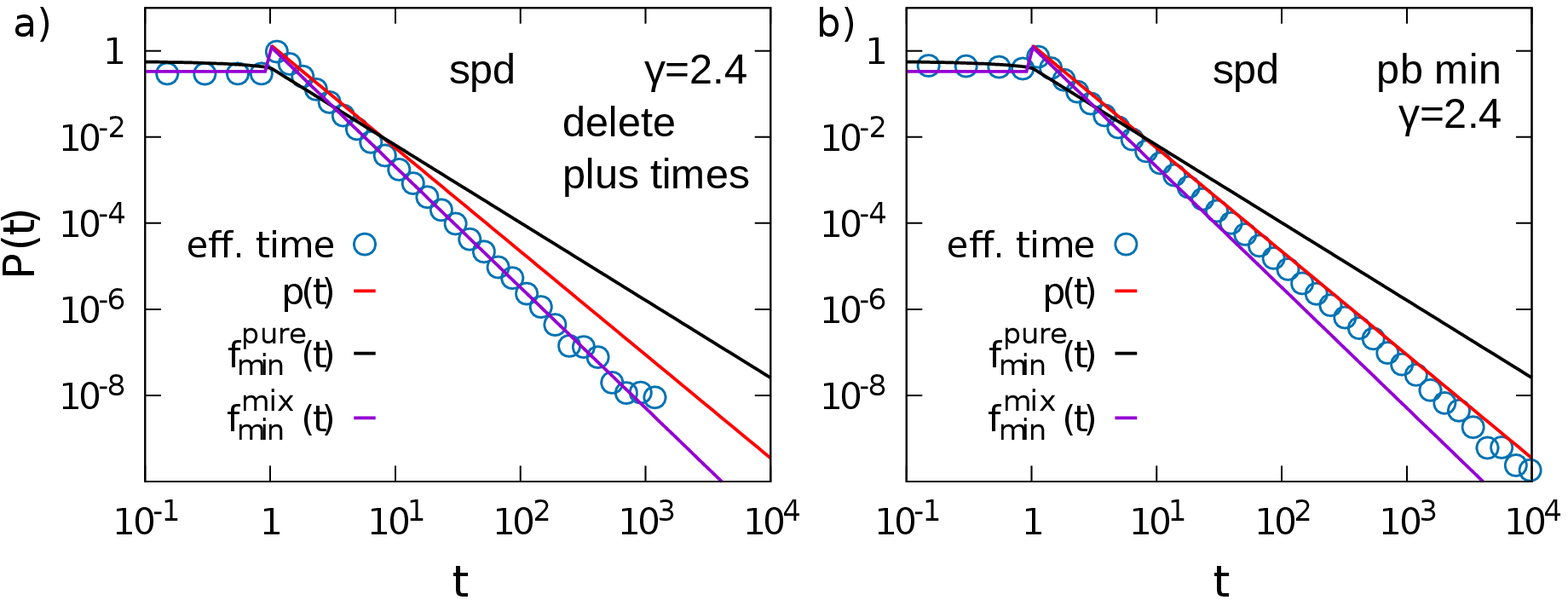}
	\caption{a) Single-particle dynamics exclusion process where residual waiting times of the plus particle is deleted after each exchange process, independent whether the plus particle was active or passive in the exchange. b) The normal spd minimum rule from the bulk text for comparison.}
	\label{fig:particle-based_min_check}
\end{figure}

In a first step, we will compare the PDF for effective waiting times of the spd minimum rule to the case where we always assign new waiting times to the plus particle after an exchange process. In figure \ref{fig:particle-based_min_check} (a), the simulation results really follow the respective estimate $f_{min}^ {mix}$, which is expected since the plus particle always has a non-aged waiting time. This result is in contrast to the result of the bulk text, which is also shown in figure \ref{fig:particle-based_min_check} (b) for comparability. For the minimum rule, effective waiting times do not follow $f_{min}^{mix}$, hence the age of the plus particle plays a role. However, neither do they follow $f_{min}^{pure}$ which is expected for the minimum of to random variables distributed by the residual time $r(t)$.

\begin{figure}[t!]
	\centering
	\includegraphics[width=.95 \textwidth]{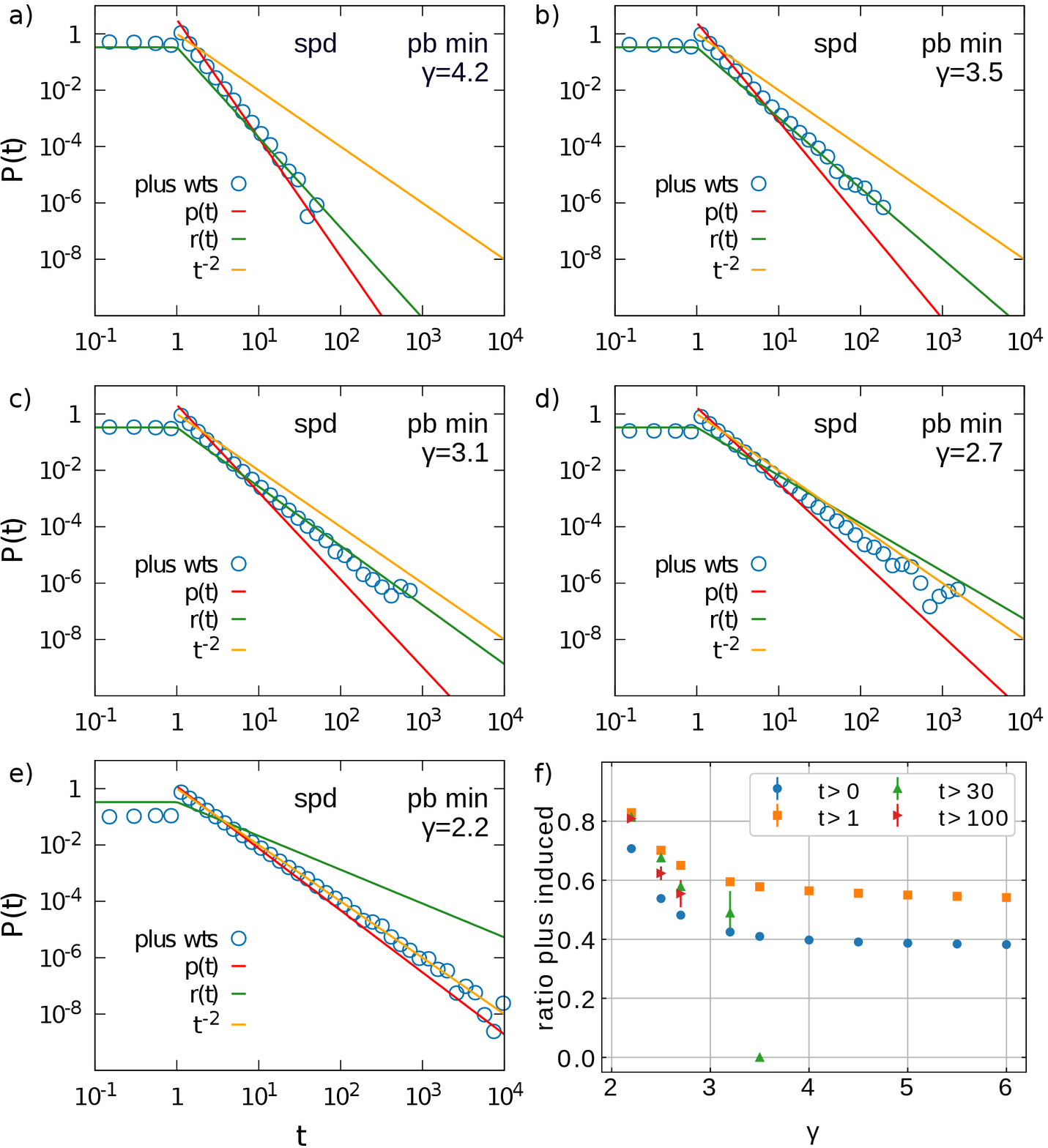}
	\caption{\textbf{Particle-based minimum rule} PDFs of the residual waiting time of a plus particle $\tilde{p}$ for a system of $L=100$ filled by $N_+=1$ and $N_-=L-1$ (spd). (a) $\gamma=4.2$, (b) $\gamma=3.5$, (c) $\gamma=3.1$, (d) $\gamma=2.7$, (e) $\gamma=2.2$. Blue data points show the residual waiting times of the plus particle, red $p(t)$ original waiting time distribution, green $r(t)$ residual waiting time from renewal theory, yellow a constant function $t^{-2}$ as a reference line. f) The measured ratio of exchange processes which have been induced by an active plus particle in the spd minimum case. Statistics over all exchange events are colored blue, exchange processes with an effective waiting time of at least 1 are orange, 30 green and at least 100 red. Data points are missing if no such high waiting times have been observed in the simulation, errors bars show the sem.}
	\label{fig:particle-based_min_plusres}
\end{figure}

In a second step, we show simulation results for residual waiting times of the plus particle in figure \ref{fig:particle-based_min_plusres}, which we call $\tilde{p}(t)$ in the following. In the fast regime of $\gamma>3$, the measurements follow the theoretical estimate $r(t)$. However, this changes for exponents $\gamma<3$. The simulation result do not follow $r(t)$ anymore but rather stay close to the asymptotic of $\gamma=2$. This slope for $\tilde{p}(t)$ is consistent with the effective exchange time observed in figure \ref{fig:particle-based_min} (a), where simulation results follow $p(t)\approx t^{-\gamma}$ in the tail, when considering the minimum out of a random variable distributed by $r(t)$ for minus particles and $\tilde{p}(t)$ for plus.


We see that the time-averaged estimate $r(t)$ is not valid anymore for the residual waiting time of a single plus particle in the minimum model for $\gamma<3$. We further give an argument for the break down of validity by showing that the fraction of exchange events induced by the plus particle increases for $\gamma<3$ (see figure \ref{fig:particle-based_min_plusres} (f)). This is in particular important for large times in the tail. If most events are induced by the single plus particle the motion is more and more determined by this particle itself and hence, the dynamics are more similar to the asymmetric particle-based model of passive minus particles.

\section{Update scheme}\label{app:update}

For the Markovian TASEP: We also remark that, thanks to the memoryless property of the Markovian TASEP, one can practically generate the next time $ t_i^{k+1} $ at every $ t= t_i^k $. 
The Markov property does not hold for the algebraic distribution $p(t)$ in equation (\ref{eq:density}). To evolve the system in time, we use a modified waiting time algorithm (\textit{next reaction method} \cite{Gibson2000}) similar to \cite{Concannon2014, Gorissen2012} .

Times for all events (particle-based or site-based) are initialized at the beginning of the simulation ($\tau=0$). The shortest time $t_\alpha$ is then chosen from a list of all waiting times to be the absolute time for the next event $\alpha$. In the realization of the event, the system time is increased up to this point in time $\tau_{i+1} = \tau_i + t_\alpha$. Anyhow, the process is only executed if the local particle configuration is appropriate. After the realization, the waiting time of the event is renewed by taking a new time $t_{new}$ from the distribution $p(t)$ added to the current system time $t_\alpha = \tau+t_{new}$. This time is then placed into the list for the event $\alpha$ and the procedure is repeated.

\newpage
\bibliography{bibliographic_database.bib}
\bibliographystyle{unsrt}

\end{document}